\documentstyle[12pt]{article}
\begin{document}
\centerline{\bf TDHFB-Langevin approach to the nuclear collective dynamics\footnote{
in Proc. Int. Symposium {\it Large-Scale Collective Motion of Atomic Nuclei}, 
October 15-19, 1996, Brolo (Messina), Italy, Edited by G. Giardina, G. Fazio, M. Lattuada, World Scientific, Singapore, 1997, p. 222. }}
\vspace{1.5cm}
\centerline{ M. Grigorescu}
\vspace{1.5cm}
\noindent
$\underline{~~~~~~~~~~~~~~~~~~~~~~~~~~~~~~~~~~~~~~~~~~~~~~~~~~~~~~~~
~~~~~~~~~~~~~~~~~~~~~~~~~~~~~~~~~~~~~~~~~~~}$ \\[.3cm]
Stochastic mean-field equations are derived
applying the time-dependent variational principle to a quantum
many-body system in interaction with a classical heat bath. 
This approach is tested on the case of a charged particle in thermal radiation field, and 
is used  to include frictional and random forces in the TDHFB  equations. Dynamical estimates of the diffusion coefficient for the quadrupole deformation are obtained. 
\\
$\underline{~~~~~~~~~~~~~~~~~~~~~~~~~~~~~~~~~~~~~~~~~~~~~~~~~~~~~~~~
~~~~~~~~~~~~~~~~~~~~~~~~~~~~~~~~~~~~~~~~~~~}$ \\[1cm]
\newpage
\section{Introduction}
The simulation of the non-equilibrium fluctuation phenomena
observed in hot nuclei represents a fundamental problem of the many-body
theory, demanding stochastic extensions of the kinetic or
mean-field equations. \\ \indent
At the level of the semiclassical description provided by the phase-space 
density, the fluctuations can be  accounted by
the Boltzmann-Langevin approach \cite{ag}. Within this
approximation the nuclear Boltzmann 
equation is modified, so that a random component 
(Brownian force \cite{ccgr}) is included in the  collision term.    
\\ \indent
The time-dependent variational calculations on  coherent states trial manifolds, as time-dependent Hartree-Fock (TDHF), or antisymmetrized molecular - dynamics 
(AMD), lead to  deterministic mean-field equations, but recipes to   
introduce a random behavior have been proposed \cite{resu, su, onra, iwa, bbb}.
The basic assumption is that some decoherence mechanism exists, interrupting 
the deterministic evolution of the quantum wave packet by
spontaneous collapse to the neighbouring states of the same trial manifold. 
The collapse rate is obtained using the Fermi's golden rule, for a suitable
residual interaction.  \\ \indent
The mechanism of the random spontaneous collapse induced by
environmental (residual) fluctuations
 remains obscure, beyond the framework
of the quantum many-body theory. This ambiguous mechanism
can be avoided if the residual interactions and the unretained degrees of 
freedom are accounted from the very beginning in the time-dependent 
variational principle. Particularly suited for this purpose is the
model of bilinear coupling between a particle and a heat bath of
harmonic oscillators, applied with success to derive the classical
Langevin equation \cite{hang}. This model can be  used also to include
stochastic forces in the mean-field equations, and   
was applied to study the effects of pairing, dissipation and 
temperature on the giant  quadrupole resonance (GQR) \cite{gc}.
\\ \indent
The dynamical approach to the coupling between a quantum many-body
system and the heat bath is presented in Sec. 2, together with a test on the
important case provided by a charged particle in blackbody radiation
field.  The stochastic mean-field equations are used in Sec. 3 
to calculate the diffusion coefficient of the quadrupole deformation
coordinate. Conclusions  are drawn in  Sec. 4. 
\section{Brownian quantum dynamics }
The phenomenology of the Brownian diffusion in  classical mechanics
can be described by two external force terms in the Hamilton equations
of motion for the particle: a random force with zero 
mean, $\xi(t)$, (the noise), and a dissipative friction force, $f$. 
The type of friction is
determined by the memory function $\Gamma(t)$, or the
spectral density of the environment, $J(\omega)$, so that
for a linear dissipation mechanism 
\begin{equation}
f (t) = - \int_0^t \Gamma(t-t') \dot{Q}(t') dt'~~, ~~
\Gamma(t) \equiv \frac{2}{\pi} \int_0^\infty d \omega 
\frac{ J(\omega) cos( \omega t) }{ \omega}~~,
\end{equation}
with $\dot{Q}$ the particle velocity.
The memory function gives also the noise correlation function, and
$\xi(t)$ should obey the fluctuation-dissipation theorem,
\begin{equation}
 << \xi(t) \xi(t') >> = k_B T \Gamma(t-t')~~, 
\end{equation}
where $<<..>>$ denotes the average over an ensemble of trajectories.
\\ \indent
Noise and friction forces having these properties may be derived 
following a Hamiltonian treatment of the whole system consisting
of the particle in interaction with 
a heat bath of $N_c$ harmonic oscillators via the bilinear coupling term
\cite{hang}
\begin{equation}
H_{coup} = Q \sum_{j=1}^{N_c} C_j q_j ~,
\end{equation}
where $C_j$ are coupling constants and $q_j$ the time-dependent bath
coordinates. \\ \indent 
The advantage of the bilinear coupling model over
the phenomenological Langevin approach becomes striking when
the particle is a quantum object. For example,  below a
certain crossover temperature, the mechanism of chemical
reactions changes from thermal activation to tunneling, becoming sensitive to 
destruction of the phase coherence and dissipation produced by
environment. The temperature effect on the tunneling rate can be 
accounted by the thermal average of the quantum penetrability factors \cite{bend}. However, 
dissipation requires a more elaborate treatment, based essentially on 
Euclidean path integral calculations for the bilinear coupling 
Hamiltonian \cite{cl}.
\\ \indent
The problem of dynamical coupling  between a quantum system and its 
environment is even more complex, having not yet a satisfactory solution.
Thus, no standard method exists to treat a mixed system composed
of a quantum particle interacting with a classical heat bath. 
In quantum gravity the similar problem of
a quantum system coupled to the classical space-time 
foam could be treated by Hamilton-Heisenberg equations derived
from a variational principle \cite{aa}. In the quantum many-body
theory,  the collective mean-field obtained from time-dependent variational calculations has 
already a hybrid character, including both quantum and classical aspects. Thus,
it is natural to define its dynamical coupling to a classical heat bath 
of oscillators using the variational equation
\begin{equation}
\delta \int dt [ \sum_{i=1}^{N_c} m_i
( \dot q_i )^2  + \langle \Psi \vert i \hbar \partial_t - 
({\bf H}_0 + {\bf H}_{b}) \vert \Psi \rangle] =0
\end{equation} 
where ${\bf H}_0$  is 
 the Hamiltonian operator for the isolated quantum system, and  ${\bf H}_b$ 
contains the bath energy plus the coupling interaction. \\ \indent
Let us consider variations of the quantum wave function  $\Psi$
in Eq. (4) restricted to a finite dimensional trial manifold 
 ${\cal S}= \{ \vert \Psi \rangle(X) \}$, parameterized by $2N$ coordinates 
 $X \equiv \{ x^i \}$, $i=1, 2N$, so that the matrix
 $ \omega^{\cal S} = [ \omega_{ij}^{\cal S} ( \Psi)]$, defined by  
\begin{equation}
\omega_{ij}^{\cal S} ( \Psi) =
2 \hbar Im \langle \partial_i \Psi \vert \partial_j \Psi \rangle 
\end{equation}
is non-singular \cite{var}. Supposing a
coupling interaction ${\bf H}_{coup} = {\bf K} \sum_{j=1}^{N_c} C_j q_j $,
linear in the coordinate operator ${\bf K}$,
the equations of motion determined by variations in Eq. (4)  
with respect to $X(t)$ and the bath trajectories $q_i (t)$ are
\begin{equation} 
 \sum_{j=1}^{2N} \dot{x}^j \omega_{jk}^{\it S}(\Psi) = 
\frac{\partial \langle \Psi \vert {\bf H}_0
\vert \Psi \rangle}{ \partial x^{k}} + \frac{ \partial \langle \Psi \vert {\bf K} 
\vert \Psi \rangle}{\partial x^k} \sum_{i=1}^{N_c} C_i q_i 
\end{equation}
for the quantum system and 
\begin{equation}
\dot q_i = \frac{p_i}{m_i}~~,~~
\dot p_i = - m_i \omega_i^2 q_i - C_i \langle \Psi \vert {\bf K} \vert \Psi \rangle
\end{equation}
for the classical oscillators. 
The classical equations can be solved in terms of the unknown function 
of time $Q_\psi=\langle \Psi \vert {\bf K} \vert \Psi \rangle$, and when their retarded solution is 
inserted in Eq. (6), a Langevin-like equation  is obtained,  
\begin{equation}
\sum_{j=1}^{2N} \dot{x}^j \omega_{jk}^{\cal S}(\Psi) = 
\frac{\partial \langle \Psi \vert {\bf H}_0 + {\bf W}_{ren}
\vert \Psi \rangle}{ \partial x^{k}} 
- \frac{\partial \langle \Psi \vert {\bf K}
\vert \Psi \rangle}{ \partial x^{k}} \lbrack \xi (t) + f_\psi \rbrack~~.
\end{equation}
Here $f_\psi$ is given by Eq. (1) with $Q_\psi$ instead of $Q$. The term 
$ {\bf W}_{ren} = {\bf K} [ Q_\psi(0) \Gamma(t) - Q_\psi(t) \Gamma(0)] $
will be neglected, supposing that it can be accounted by renormalizing ${\bf H}_0$. \\ \indent
If ${\cal S}$ is the whole Hilbert space of the quantum system,
then Eq. (8)  becomes a non-linear time-dependent Schr\"odinger equation 
\begin{equation}
i \hbar \frac{ \partial \Psi}{ \partial t} = [{\bf H}_0 
 - {\bf K}(\xi(t) + f_\psi ) ] \Psi ~~.
\end{equation}
The non-linearity appears in the friction term $f_\psi$,
accounting for
the backreaction of the environment on the quantum system. When friction 
 is neglected,  the rate of the noise-induced excitation 
between two eigenstates $\vert E_i\rangle$ and  $ \vert E_f\rangle$ of ${\bf
H}_0$ can be easily calculated
in the first-order approximation for the time-dependent
perturbation represented by ${\bf H}_{noise} = - {\bf K}
\xi(t)$. The result is  
\begin{equation}
\Gamma_{fi} = \frac{ \vert {\bf K}_{fi} \vert^2 }{ \hbar^2} lim_{t
\rightarrow \infty} \frac{1}{t}
\vert \int_0^t \   \xi(s) e^{ i \omega_{fi} s } ds \vert^2
\end{equation}
and can be further averaged over the statistical ensemble before the time-integral making use of Eq. (2), so that
\begin{equation}
\Gamma_{fi}^{noise} \equiv << \Gamma_{fi} >> =
   \frac{2  }{\hbar^2} \vert {\bf K}_{fi} \vert^2 k_B T
\frac{ J(\omega_{fi})}{\omega_{fi}} ~~.
\end{equation}
Here ${\bf K}_{fi} = \langle E_f \vert {\bf K} \vert E_i \rangle $ is the matrix
element of the coupling operator ${\bf K}$, and $\omega_{fi}= \vert E_f - E_i \vert/
\hbar$. \\     \indent
The rate provided by Eq. (11) may be used to test the formalism 
presented above on the case of a single charged particle in thermal radiation 
field, considered as a classical heat bath of oscillators continuously
distributed in frequency. 
  In the dipolar approximation, the coupling of the electromagnetic
field to the  charge $e$  has the form ${\bf H}_{noise}^{rad} =
- \vec{{\bf d}} \cdot \vec{E}=-{\bf d}_x E_x - {\bf d}_yE_y-{\bf d}_zE_z$ where 
$\vec{{\bf d}}= e \vec{{\bf r}}$ is the electric dipole
operator, and $\vec{E}$ the radiation electric field. For isotropic
radiation  ${\bf H}_{noise}^{rad}$ is a sum of three terms  of the form 
$-{\bf K} \xi$ assumed above, each 
with the spectral density $J^{rad}(\omega) = 2 \hbar e^2 \omega^4
\langle n \rangle_\omega / (3 c^3 k_B T)$. The total transition rate $\Gamma_{fi}^{rad}$,
is the sum of three terms provided by Eq. (11), and the result
\begin{equation}
\Gamma_{fi}^{rad}
 =   \frac{4 e^2}{3 \hbar c^3    }
 \vert \vec{ {\bf r}}_{fi} \vert^2   \omega^3_{fi}   \langle n \rangle_{\omega_{fi}}, ~~~
 \langle n \rangle_\omega= {1}/{(  e^{ \hbar \omega / k_B T} -1)},
\end{equation}
coincides with the known transition rate in
thermal  radiation field \cite{pamd}.
At high temperatures  the memory function corresponding to $J^{rad}(\omega)$
can be expressed in terms of the second derivative of the delta function,
$\Gamma(t) = - 4 e^2 \ddot{\delta}(t)/(3c^3)$,  and the related friction force
\begin{equation}
\vec{f}^{rad}_\psi = \frac{2 e^2}{3c^3} \frac{d^3 \langle {\vec{{\bf r} }}\rangle}{dt^3}
\end{equation}
is the classical radiation reaction in vacuum \cite{jack}. 
\\ \indent
In the case of  Ohmic dissipation $ \Gamma (t) = 2  \gamma_K \delta (t)$, with
$\gamma_K$ the static friction coefficient, and 
Eq. (8) becomes
$$
\sum_{j=1}^{2N} \dot{x}^j \omega_{jk}^{\cal S}(\Psi) = 
\frac{\partial \langle \Psi \vert {\bf H}_0 \vert \Psi \rangle}{ \partial x^{k}} 
$$
\begin{equation}
- \frac{\partial \langle \Psi \vert {\bf K} \vert \Psi \rangle}{ \partial x^{k}} \lbrack \xi (t) 
- \gamma_K  \frac{ d \langle \Psi \vert {\bf K} \vert \Psi\rangle }{dt} 
 \rbrack.
\end{equation}
\section{ TDHFB-Langevin dynamics and the shape diffusion coefficient} 
An important application of the stochastic mean-field techniques concerns
the calculation of the shape diffusion coefficient. The thermal 
equilibrium state of a highly excited nucleus can be represented
\cite{bbb} by an  ensemble of incoherent self-consistent  
Hartre-Fock configurations $\vert \Psi_k\rangle$, with energies
$E_k=\langle{\bf H}_0\rangle_{\Psi_k}$ and quadrupole deformations
\begin{equation}
\beta_k = \frac{ 4 \pi}{5} 
\frac{ \langle{\bf Q}_0^p+{\bf Q}_0^n\rangle_{\Psi_k}}{A \langle r^2 \rangle}
~~,~~{\bf Q}_0^\nu = \sum_{i=1}^{N_\nu} \sqrt{\frac{5}{16 \pi}} (2 {\bf z}^2 -{\bf x}^2 -
{\bf y}^2)_i^\nu ~~.
\end{equation}
Within this ensemble the transitions induced by the residual interactions
occur with a rate 
\begin{equation}
\Gamma_{fi} = \frac{2 \pi}{\hbar}  \vert \langle \Psi_f \vert {\bf H}_{res} \vert 
\Psi_i\rangle
\vert^2 \delta(E_f-E_i)
\end{equation}
and a diffusion coefficient of the quadrupole deformation can be defined by \cite{bbb} 
\begin{equation}
D_\beta= \sum_f (\beta_f-\beta_i)^2 \Gamma_{fi}~~,
\end{equation}
where the sum involves states $\vert \Psi_f\rangle$ different by $\vert \Psi_i\rangle$ in the occupation number of two orbital levels. However, the value
predicted by this calculation is only a fraction $(\approx 10\%)$ of the 
estimate provided by the Einstein formula\footnote{When $x_t$ is a classical Brownian trajectory,  $lim_{t \rightarrow \infty} << (x_t -x_0)^2>> /t =2 k_B T/ \gamma_x $.}
\begin{equation}
D_\beta^E = \frac{k_B T}{\gamma_{\beta}}
\end{equation}
with the friction coefficient $\gamma_\beta$ extracted from experiment. 
\\ \indent
A dynamical calculation of $D_\beta$ can be performed
using  an ensemble of Brownian trajectories $\beta(t)$ generated
by Eq. (14) for a nucleon system  in contact with a heat bath. 
The study of the axially symmetric quadrupole shape vibrations in the sd shell nucleus
$^{28}$Si including pairing, friction and temperature 
shows that  the occupation numbers $p_r, r=1,6 $ 
of the orbital levels depend on time and may switch between different 
configurations, but each configuration is quasi-stable for relatively long 
periods, despite the shape fluctuations.  For 
$^{28}$Si  this switching means a 
change of $\{p_r \}$ from the oblate configuration
of the metastable ground state (mgs), $\{ p_r^0 \}$, to an unstable prolate 
configuration, $\{ p_r^* \}$. The rate of these oblate$\rightarrow$prolate
configuration transitions can be defined in terms of the
mean-first-passage-time  \cite{hang},  $\tau^*$, which is
the average time elapsed until the system switches the configuration
for the first time. Therefore, a diffusion coefficient of the (quasi-stable) quadrupole deformation  can be defined  by the average
\begin{equation}
D_\beta^* = \frac{1}{N_b} \sum_{b} \frac{(\beta_i^b -\beta_f^b)^2}{\tau^*_b}~~~,~~~
\beta_i^b \equiv \beta^b(0)~~,~~\beta_f^b \equiv \beta^b(\tau^*_b) 
\end{equation}
where the sum is taken over $N_b$ trajectories generated with random initial 
conditions at a fixed energy $E_i$.
For such calculations, the trial manifold ${\cal S}$ in Eq. (4),
which is the best suited to account for the occupation number degrees of freedom, 
consists of a set ${\cal S}^{HFB}$ of normalized Hartree-Fock-Bogolyubov functions. In this case, without environment coupling, Eq. (14) corresponds to
the deterministic time-dependent Hartree-Fock-Bogolyubov (TDHFB) equations.
The sd shell  contains $N_s=12$ states, and ${\cal S}^{HFB}$ is parameterized
by  $N_s(N_s-1)=132$  variables $\{ x^i \}$.
 This number is large, and therefore it is convenient to restrict
${\cal S}^{HFB}$ to a  submanifold with lower dimensionality,
accounting only for the quadrupole and occupation numbers degrees of freedom.  
For the 6 protons and 6 neutrons of $^{28}$Si outside the
$^{16}$O core, the 132-dimensional manifold ${\cal S}^{HFB}$
can be reduced by the cranking technique \cite{pred95}
to a 14-dimensional 
submanifold  corresponding to 7 degrees of freedom: 
$\beta$ and  the occupation numbers $p_r, r=1,6$ of the sd shell
orbital levels.
\\ \indent  
The  model Hamiltonian ${\bf H}_0$ includes  
the spherical harmonic oscillator one-body term ${\bf h}_{0}$ and the  
quadrupole-quadrupole plus pairing two-body interactions:
\begin{equation}
{\bf H}_0= {\bf h}_{0} - \chi {\bf K}^{2} - \frac{G_{pair}}{4} 
{\bf P}^{\dagger} {\bf P} ~~.
\end{equation}
Here ${\bf K} \equiv {\bf Q}_0 / c_0$,  $c_0 = (5/ 4 \pi)^{1/2} \hbar / m 
\omega_0 = 26.17$MeV$\cdot$fm$^2/ \hbar \omega_0$, 
$\omega_0= 41 A^{-1/3}$ MeV/$\hbar$, and  the interaction strengths
are chosen to be
$\chi = 0.186$ MeV and $G_{pair}=1.23$ MeV.
The  mgs configuration ($\beta^0,p_r^0$) corresponds to an oblate shape
($\beta^0 =-0.43$), and  is obtained by frictional 
cooling \cite{hor}. This  procedure ensures the
self-consistency of the  quadrupole and  pairing mean-fields \cite{gc}.
\\ \indent
 The operator ${\bf K}$ is also the  coordinate appearing in the 
coupling to the heat bath. The corresponding friction
coefficient  is related to $\gamma_\beta$ by 
$\gamma_K  = 0.0083 \gamma_\beta$, and at $T=0$ the rate $\Gamma$
of the energy dissipation   is the same as the decay
width of the GQR, $\Gamma_{GQR} \approx
4.5 $ MeV,  when $\gamma_\beta \sim 6.4 \hbar$. \\ \indent
The average in Eq. (19) was calculated using 300 trajectories starting at the GQR
energy of  $19.1$ MeV,  with random initial conditions. When 
$k_BT$ is between 1.5 and 2.5 MeV, and $\gamma_\beta$ is near
$6.4 \hbar$, then  $D_\beta^*$   
can be  accurately interpolated by the formula
\begin{equation}
D_\beta^* (T) = \frac{ k_B T}{u + v k_B T+ \gamma_\beta( a + b k_B T)}
\end{equation}
where $u=0.27 \hbar$, $v= 1.1 \hbar /$MeV, 
$a=0.14 $ and $b=0.1 $MeV$^{-1}$. This result indicates that
the effective friction coefficient for shape diffusion with configuration 
change,  $\gamma^*_\beta = u + vk_BT + \gamma_\beta( a + b k_B T)$ 
is not the same as the fixed model parameter $\gamma_\beta$,  but depends on 
temperature. For $k_BT= 2.5$ MeV, Eq. (21) predicts
a diffusion coefficient $D_\beta^* = 453$ keV$/ \hbar$, about a factor ten
larger than the microscopic estimate \cite{bbb} of  $30$
keV$/ \hbar $ in the neighboring nucleus $^{24}$Mg. Though, assuming an $A^{-1}$ dependence of $D_\beta^*$ on the mass number, for $^{158}$Er one obtains
$D_\beta^* \sim 80$ keV$/ \hbar$, which is close to the value of $\sim 70$ 
keV$/ \hbar$ extracted  from experimental data. 
\section{Conclusions}   
The variational principle applied to a quantum many-body system and its thermal environment provides stochastic mean-field equations (Eq. (8)). Depending on the trial manifold,
 a non-linear Schr\"odinger-Langevin equation (Eq. (9)), or a system of TDHFB-Langevin equations can be obtained.  The example of a charged particle in blackbody
radiation field shows that this formalism predicts correctly the 
transition rate  and radiation reaction.  
\\ \indent
The stochastic mean-field equations (TDHFB-Langevin)
have been applied to calculate the diffusion coefficient of the 
quadrupole deformation (Eq. (21)). The effective friction coefficient 
$\gamma^*_\beta $   for shape diffusion with change of the 
internal  configuration is not the same as the model parameter $\gamma_\beta$
and depends on temperature.
The results of the present calculations within the sd shell 
are consistent with the measured values of the  GQR decay width
and shape diffusion coefficient,
proving the relevance  of the TDHFB-Langevin formalism.

\end{document}